\documentclass[english,prb,preprint,superscriptaddress,showpacs,preprintnumbers,amsmath,amssymb]{revtex4}

\usepackage{graphicx}
\usepackage{float}
\usepackage{dcolumn}
\usepackage{bm}
\usepackage{epsfig}
\newcommand{\ignore}[1]{}

\begin{document}

\title{A first principles study on the electronic and magnetic properties of Ba$_{1-x}$K$_x$Fe$_2$As$_2$}

\author{Jun Dai}
\affiliation{Hefei National Laboratory for Physical Sciences at
Microscale, University of Science and Technology of China, Hefei,
Anhui 230026, P.R. China}

\author{Zhenyu Li}
\affiliation{Hefei National Laboratory for Physical Sciences at
Microscale, University of Science and Technology of China, Hefei,
Anhui 230026, P.R. China}

\author{Jinlong Yang}
\thanks{Corresponding author. E-mail: jlyang@ustc.edu.cn}
\affiliation{Hefei National Laboratory for Physical Sciences at
Microscale, University of Science and Technology of China, Hefei,
Anhui 230026, P.R. China}

\author{J. G. Hou}
\affiliation{Hefei National Laboratory for Physical Sciences at
Microscale, University of Science and Technology of China, Hefei,
Anhui 230026, P.R. China}

\date{\today}

\begin{abstract}
We report a systematic first-principles study on the recent
discovered superconducting Ba$_{1-x}$K$_x$Fe$_2$As$_2$ systems ($x$
= 0.00, 0.25, 0.50, 0.75, and 1.00). Previous theoretical studies
strongly overestimated the magnetic moment on Fe of the parent
compound BaFe$_2$As$_2$. Using a negative on-site energy $U$, we
obtain a magnetic moment 0.83 $\mu_B$ per Fe, which agrees well with
the experimental value (0.87 $\mu_B$). K doping tends to increase
the density of states at fermi level. The magnetic instability is
enhanced with light doping, and is then weaken by increasing the
doping level. The energetics for the different K doping sites are
also discussed.


\end{abstract}


\maketitle

\section{INTRODUCTION}
The recent discovery of superconductivity in LaFeAs[O,F] has
intrigued tremendous interest in layered FeAs systems.\cite{jacs}
Intensive studies have revealed that, by substituting La with Ce,
Sm, Nd, Pr, and Gd, \cite{syn-Ce,syn-Sm,syn-Nd,syn-Pr,syn-Gd} the
superconducting temperature ($T_c$) can be raised from 26 up to 53.3
K, and even higher (about 55 K) under high
pressure.\cite{nature,syn-Sm-hp} As we know, the parent compound of
the these superconductors has a tetrahedral ZrCuSiAs-type structure
with alternate stacking of tetrahedral FeAs layers and tetrahedral
LaO layers, and favors a stripe like antiferromagnetic (AFM) ground
state. The parent compound is not a superconductor but a poor metal
with high density of states and low carrier density. \cite{lda1} The
ground state of the parent compound is supposed to be a spin density
wave (SDW) ordered state with a stripe like AFM configuration.
\cite{lda6, ns1} Superconducting occurs when the SDW instability is
suppressed by replacing of O with F  or importing O vacancies
(electron doping), or Sr substituting of La (hole
doping).\cite{syn-Sm,vacancy,epl}

Very recently, the family of FeAs-based supercondutors has been
extended to double layered RFe$_2$As$_2$ (R=Sr,Ba,Ca).
\cite{Ba1,Sr1,Sr2,Ca1,Ca2,Ca3} The electronic structure of the
parent compound has been studied both experimentally
\cite{Ba2,Ba3,Sr3} and theoretically. \cite{Ba4,LDA2,LDA3} The
density of states of RFe$_2$As$_2$ is very similar to that of
ReFeAsO around the fermi level, so does the fermi surface. The
magnetic order of BaFe$_2$As$_2$ has been revealed by
experiment,\cite{Ba5} and the magnetic moment on Fe is 0.87 $\mu_B$.
Besides, SDW anomaly has also been found in the RFe$_2$As$_2$
systems.\cite{Ba2}

Although the superconducting mechanism of these new superconductors
is still unclear, the peculiar properties of the FeAs layers,
especially the magnetic properties, are believed to be very
important for understanding the origin of the superconductivity in
these compounds. Although theoretical works have been reported for
the double layered FeAs superconductors, the doping structure,
magnetic coupling, as well as the the electronic structure after
doping have not been thoroughly investigated. Besides, the magnetic
moment on Fe atom obtained from previous theoretical studies is much
larger than the experimental value (cal. 2.67 $\mu_B$ v.s. exp. 0.87
$\mu_B$). \cite{Ba4,Ba5} Similar problem has been encountered for
the single layered ReFeAsO superconductors, and it was suggested
that a negative on-site energy $U$ should be applied to such
systems. \cite{neg-U} It is interesting to see if such a remedy also
works for BaFe$_2$As$_2$. Although the use of a negative U is
counterintuitive, it is physically possible. As suggested in a very
recent work, \cite{neg-U2} in itinerant systems, for d$^6$
configuration as Fe$^{2+}$ is, the exchange-correlation effect may
cause charge disproportionation (2d$^6$ $\to$ $d^5+d^7$) and lead to
$U=E(N+1)+E(N-1)-2E(N)<0$.

In this paper, we report the theoretical electronic and magnetic
properties of Ba$_{1-x}$K$_x$Fe$_2$As$_2$ ($x$ = 0.00, 0.25, 0.50,
0.75, and 1.00) from first-principles calculations in the framework
of generalized gradient approximation(GGA)+U. With a negative $U$,
we obtain a magnetic moment per Fe atom for BaFe$_2$As$_2$ equal to
0.83 $\mu_B$. By comparing the total energies, we predict the most
favorable doping structure. Moreover, we find slight doping ($x$
near or small than 0.25) tends to enhance the magnetic instability,
while medium and heavy dopings ($x$ near or larger than 0.5) tend to
suppress it.

\section{MODEL AND METHOD}
\begin{figure}
\includegraphics[width=8.5cm]{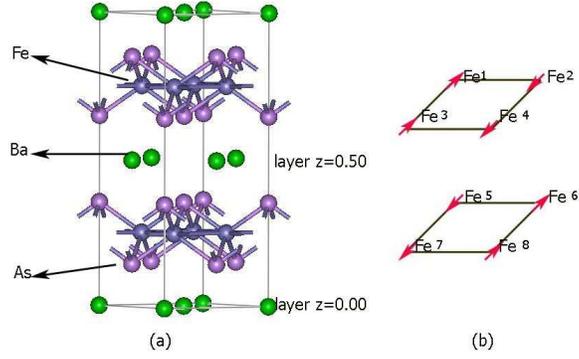}
\caption{(Color online) (a) The crystal structure of the
$\sqrt{2}\times\sqrt{2}\times1$ BaFe$_2$As$_2$ supercell. (b) The
two Fe planes in the supercell. Red arrows show the AFM4
configuration.} \label{fig1}
\end{figure}

BaFe$_2$As$_2$ exhibits the ThCr$_2$Si$_2$-type structure (space
group $I4/mmm$), where FeAs layers are separated by single Ba layers
along the c axis as shown in fig.\ref{fig1} (a). The FeAs layers are
formed by edge-shared FeAs$_4$ tetrahedra, similar to that in
ReFeAsO. In the calculation, we adopt a
$\sqrt{2}\times\sqrt{2}\times1$ supercell, which contains four Ba
atoms, eight Fe atoms, and eight As atoms. All structures are fully
optimized until force on each atom is smaller than 0.01 eV/\AA.
During all the optimizations and static calculations, the lattice
parameters are fixed to the experimental values $a=b=5.53$ \AA\/ and
$c=13.21$ \AA.\cite{Ba1} Although the lattice constants are
different at different doping levels, the variations are very small,
and we think they will not significantly change the electronic
structures of the systems. To simulate doping, we replace one, two,
three, and four Ba atoms with K atoms, which corresponds to 25\%,
50\%, 75\%, and 100\% doping, respectively.

The electronic structure calculations are carried out using the
Vienna \textit{ab initio} simulation package\cite{vasp} within
GGA+U.\cite{GGA} The electron-ion interactions are described in the
framework of the projected augment waves method and the frozen core
approximation.\cite{PAW} The energy cutoff is set to 400 eV. For
density of states (DOS) calculation, we use a 12$\times$12$\times$6
Monkhorst dense grid to sample the Brillouin zone, while for
geometry optimization , a 6$\times$6$\times$3 Monkhorst grid have
been used.  The on-site Coulomb repulsion is treated approximately
within a rotationally invariant approach, so only an effective U,
defined as $U_{eff}$=U--J needs to be determined, where U is the
on-site Coulomb repulsion (Hubbard U) and J is the atomic-orbital
intra-exchange energy (Hund's parameter)\cite{ldau}. Here we adopt a
negative $U_{eff}$ of -0.5 eV, and if not specially mentioned, all
the discussions in the results are based on $U_{eff}= -0.5$ eV.

\begin{figure}
\includegraphics[width=8.5cm]{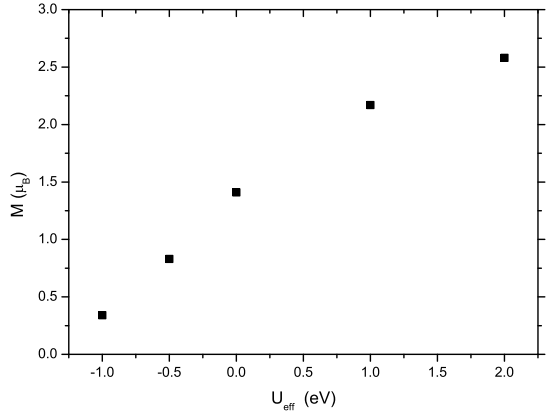}
\caption{The calculated magnetic moment of the AFM4 state of the
$\sqrt{2}\times\sqrt{2}\times1$ BaFe$_2$As$_2$ with different
U$_{eff}$.} \label{fig2}
\end{figure}

\begin{figure}
\includegraphics[width=8.5cm]{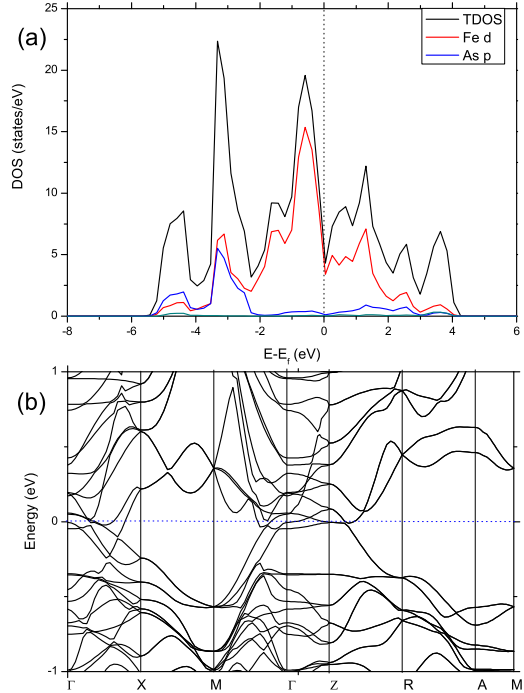}
\caption{(Color online) The total DOS and the projected DOS of the
Fe d states and the As p states of the AFM4 state of the
$\sqrt{2}\times\sqrt{2}\times1$ BaFe$_2$As$_2$. Since the spin-up
and spin-down states are degenerated for AFM states, we plot the
spin-up channel only.} \label{fig3}
\end{figure}

\section{RESULTS AND DISCUSSIONS}
First, we focus on the electronic properties of the mother compound
BaFe$_2$As$_2$. In order to describe the electronic structures with
different magnetic orderings, the Fe atoms in two planes are
numbered as in fig.\ref{fig1} (b). Except for the nonmagnetic(NM)
and ferromagnetic (FM) states, the system have six possible AFM
states: square-like in-plane AFM with Fe atoms directly above each
other in the c-direction aligned parallelly, AFM1
(--,--,+,+,--,--,+,+),  and antiparallelly, AFM2
(--,--,+,+,+,+,--,--); stripe-like in-plane AFM with Fe atoms
directly above each other in the c-direction aligned parallelly AFM3
(--,+,--,+,+,--,+,--), and antiparallelly AFM4
(--,+,--,+,--,+,--,+,--,+); one plane with square-like AFM and the
other with stripe-like AFM, AFM5 (--,--,+,+,+,--,+,--); and in-plane
FM with two planes aligned antiparallelly, AFM6
(--,--,--,--,+,+,+,+). We initialize the systems with these NM, FM
and six AFM orderings. After SCF calculations, the AFM1, AFM2, and
AFM3 states converge to the NM state. The instabilities of NM state
to other magnetic states, and the corresponding magnetic moment of
Fe in these magnetic states are listed in Table \ref{Table1} and
\ref{Table2}. We find very weak instabilities from NM to FM and
AFM6, a stronger one to AFM5, and the strongest instability to AFM4,
which is the ground state. This ground state is consistent with the
previous experimental result\cite{Ba5} and other
calculations\cite{Ba7}, where the ground state of BaFe$_2$As$_2$ was
found to be stripe-like AFM with Fe atoms aligned antiparallelly to
each other in c-direction. The magnetic moment we obtained for the
ground state is about 0.83 $\mu_B$/Fe, comparing with that in other
calculations (about 2.67 $\mu_B$/Fe), our result agrees much better
with the experimental one (0.87 $\mu_B$/Fe).

\begin{table*}
\caption{The calculated magnetic instabilities of
Ba$_{1-x}$K$_x$Fe$_2$As$_2$ (x=0.00, 0.25, 0.50 and 1.00). In the
cases of same doping level, the lowest energy among the NM states is
set to zero, energy with other magnetic configuration is the
difference to it. ($E-E_{NM}$). The energy unit is meV. }
\label{Table1}
\begin{tabular}{|c|ccccccccc|}
\hline \hline &&NM&FM&AFM1&AFM2&AFM3&AFM4&AFM5&AFM6 \\ \hline
BaFe$_2$As$_2$&&0.00&-0.89&--&--&--&-72.65&-37.04&-0.06 \\ \hline
Ba$_{0.75}$K$_{0.25}$Fe$_2$As$_2$&case
1&0.02&--&--&--&--&-91.49&-45.42&-- \\
&case 2&0.00&--&--&--&-95.40&-91.12&-45.26&-- \\ \hline
Ba$_{0.50}$K$_{0.50}$Fe$_2$As$_2$&case
1&161.48&157.92&--&161.94&--&97.55&130.38&151.23\\
&case 2&161.20&157.47&--&164.89&--&96.22&129.94&150.69\\
&case 3&0.00&-18.32&-0.28&0.33&-88.92&-88.86&-44.01&-21.03 \\ \hline
Ba$_{0.25}$K$_{0.75}$Fe$_2$As$_2$&case
1&0.00&-42.20&-9.53&-14.82&--&-54.21&-36.20&-33.00 \\
&case 2&0.00&-42.21&-9.53&-14.84&-57.86&-54.26&-36.23&-33.01 \\
\hline KFe$_2$As$_2$&&0.00&-7.15&-30.30&--&--&1.97&-15.84&-39.20 \\
\hline \hline
\end{tabular}
\end{table*}

\begin{table*}
\caption{The calculated magnetic moments in $\mu_B$ per Fe atoms of
Ba$_{1-x}$K$_x$Fe$_2$As$_2$ (x=0.00, 0.25, 0.50 and 1.00).}
\label{Table2}
\begin{tabular}{|c|cccccccc|}
\hline \hline &&FM&AFM1&AFM2&AFM3&AFM4&AFM5&AFM6 \\ \hline
BaFe$_2$As$_2$&&0.06&--&--&--&0.83&0.02, 0.85&0.03 \\ \hline
Ba$_{0.75}$K$_{0.25}$Fe$_2$As$_2$&case
1&--&--&--&--&0.81&0.01, 0.81&-- \\
&case 2&--&--&--&0.82&0.81&0.01, 0.81&-- \\ \hline
Ba$_{0.50}$K$_{0.50}$Fe$_2$As$_2$&case
1&0.20&--&0.12&--&0.78&0.35, 0.76&0.21\\
&case 2&0.20&--&0.33, 0.35&--&0.78&0.14, 0.77&0.21\\
&case 3&0.25&0.13&0.51&0.83&0.84&0.39, 0.83&0.23 \\ \hline
Ba$_{0.25}$K$_{0.75}$Fe$_2$As$_2$&case
1&0.31&0.57&0.63&--&0.78&0.80, 0.84&0.30 \\
&case 2&0.31&0.57&0.63&0.80&0.79&0.79, 0.84&0.30 \\
\hline KFe$_2$As$_2$&&0.44&1.10&--&--&0.61&0.58, 0.74, 1.04, 1.06&0.30 \\
\hline \hline
\end{tabular}
\end{table*}

We have tested the effects of different U$_{eff}$ on the magnetic
moments and total energies. As shown in fig. \ref{fig2}, the
magnetic moment changes monotonously with U$_{eff}$, and a slight
change of the U$_{eff}$ will significantly alter the magnetic
moment. Similar results have been found in ReFeAsO, \cite{neg-U} a
negative effective U thus may be a common feature of the FeAs
layers.

\begin{figure}
\includegraphics[width=8.5cm]{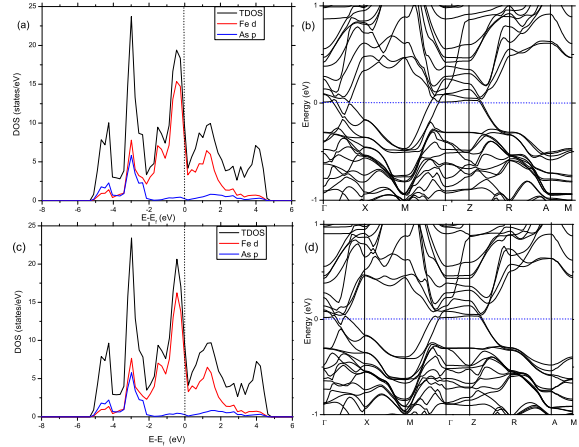}
\caption{(Color online) (a) DOS  and (b) band structure of the AFM4
state of case 1, (c) DOS and (d) band structure of the AFM3 state of
case 2 for Ba$_{0.75}$K$_{0.25}$Fe$_2$As$_2$. } \label{fig4}
\end{figure}

The density of states (DOS) of the AFM4 state is shown in fig.
\ref{fig3}a, similar to that in ReFeAsO, the contributions from Fe
and As dominate the DOS near the fermi level, and the density of
states at the fermi level (N$_{E_f}$) is 5.65. The band structure of
AFM4 is illustrated in fig. \ref{fig3}b, the small dispersions along
c axis (from $\Gamma$ to Z and A to M) indicate the interactions
between layers are small. There are three bands cross the fermi
level, one electron band around $\Gamma$ to X, and two hole bands
around $\Gamma$ to Z and M to $\Gamma$, which indicates the
multi-band feature of the system.
\begin{figure}
\includegraphics[width=8.5cm]{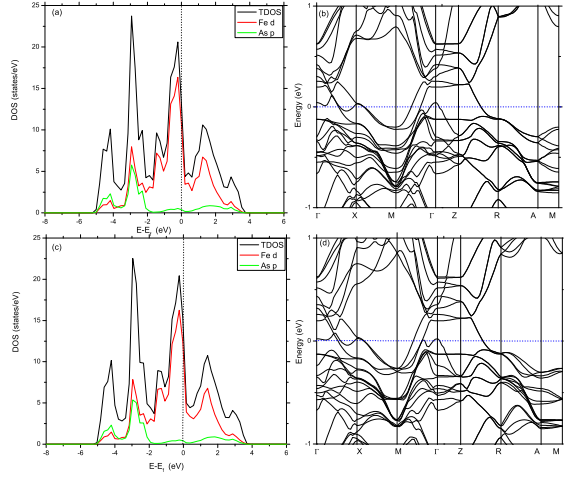}
\caption{(Color online)(a) DOS  and (b) band structure  of the  AFM3
state, (c) DOS and (d) band structure of the AFM4 state for
Ba$_{0.5}$K$_{0.5}$Fe$_2$As$_2$.} \label{fig5}
\end{figure}

Next, we turn to the doping effects on the electronic structure of
the system. In the case of one K replace of Ba
(Ba$_{0.75}$K$_{0.25}$Fe$_2$As$_2$), the K site has two choices, one
is in the layer of z=0.0 (case1), and the other is in the layer of
z=0.5 (case2), where z is the direct coordinate along the c axis of
the supercell. The total energies of these two cases are very close,
for NM state, the total energy of case 1 is about 0.2 meV higher
than that of case2. The results of magnetic instabilities and the
magnetic moment on Fe of these two cases are listed in Table
\ref{Table1} and \ref{Table2}. We find the AFM3 state disappears in
case1, while in case2, it is the state has the lowest energy, and
the magnetic moment on Fe is not significantly changed in the states
with the in-plane stripe-like AFM.

The DOS and band structures of AFM4 of case1 and AFM3 of case2 are
illustrated in fig. \ref{fig4}. Compared with the parent compound,
the shape of the DOS does not alter significantly, but the states
near the fermi level are shifted up, resulted in an increase of the
N$_{E_f}$, which is 9.46 in AFM4 of case1, and 9.18 in AFM3 of
case2.  In the band structures, the changes of the states near the
fermi level is much clearer, in both AFM4 of case1 and AFM3 of
case2, only two hole bands across the fermi level, this is accord
with the experimental results where hole pockets at $\Gamma$ sites
become larger after doping.\cite{Ba6}

\begin{figure}
\includegraphics[width=8.5cm]{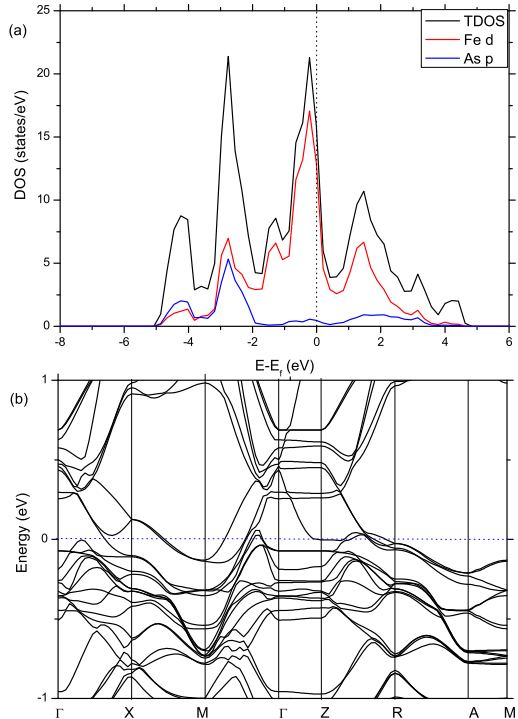}
\caption{(Color online) (a) DOS  and (b) band structure of the AFM3
state of Ba$_{0.25}$K$_{0.75}$Fe$_2$As$_2$.} \label{fig6}
\end{figure}

Then, we go to the case of Ba$_{0.5}$K$_{0.5}$Fe$_2$As$_2$. There
are three possible ways of substitution for K, that is two in the
layer z=0.0 (case 1), two in the layer z=0.5 (case 2) and one in
each layer (case 3). The calculations show case 3 is the most
favorable in the energy point of view, for NM state, the total
energy of case 3 is about 0.19 eV lower than that of case 1 and case
2 per supercell. Thus, here we discuss the magnetic and electronic
properties of case3 only. As shown in Table \ref{Table1} and
\ref{Table2}, although AFM3 is the state with the lowest energy,
AFM4 is very close to it, and the magnetic moments on Fe of these
two states are almost the same. So AFM3 and AFM4 may co-exist at
this doping level, competing with each other. The DOS and band
structure of AFM3 and AFM4 are given in fig. \ref{fig5}, though the
shape of the DOS is similar to former cases, the states are further
slightly shifted up with N$_{E_f}$ of 12.84 for AFM3 and 13.13 for
AFM4. In the band structure, there are 3 bands across the fermi
level.

\begin{figure}
\includegraphics[width=8.5cm]{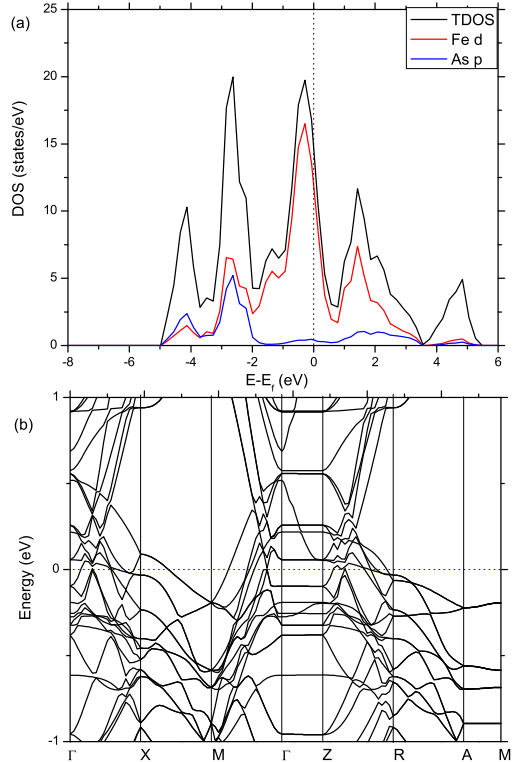}
\caption{(Color online) (a) DOS  and (b) band structure  of the AFM6
state for KFe$_2$As$_2$. } \label{fig7}
\end{figure}

For Ba$_{0.25}$K$_{0.75}$Fe$_2$As$_2$, K atoms have two choices, one
is two atoms in layer z=0.0, and the other one in layer z=0.5 (case
1), the other choice is two in layer z=0.5, and the other one in
layer z=0.0 (case 2). In spin-unpolarized calculations, these two
cases have almost the same total energy (case 1 lower about 3.65 meV
than case 2), so these two structures may co-exist at this doping
level. Although with very close energy, their magnetic structures
are different. In Table \ref{Table1}, we can see the AFM3 state is
not exist in case1, while in case2, it has the lowest energy. And
again, we find the magnetic moments on Fe of the states with the
in-plane stripe-like AFM are almost the same. Besides, at this
doping level, we find the instability from NM to FM is increased,
close to that of NM to AFM4 or AFM3. So here, the system may have
competing orders of FM, AFM3, and AFM4. In all states of case 1 and
case 2, the AFM3 of case 2 has the lowest energy, the DOS and band
structure of this state are plotted in fig. \ref{fig6}. The
N$_{E_f}$ in this case is increased to 15.24, and the bands near the
fermi level are moved up, exhibiting 5 bands across the fermi level.

Lastly, although the 100\% percent doping is hard to achieve in
experiments, we still investigate this KFe$_2$As$_2$ case for
consistency. The magnetic instabilities and magnetic moments on Fe
atoms are given in Table \ref{Table1} and \ref{Table2}, we find the
NM to AFM6 has the strongest instability here, with AFM1 the next.
This is different with the above cases, where the states with
in-plane stripe-like AFM is always the state with the lowest energy.
The DOS and band structure are shown in fig. \ref{fig7}. Comparing
with the DOS of Ba$_{0.25}$K$_{0.75}$Fe$_2$As$_2$, the N$_{E_f}$ is
slightly decreased to 14.70, and 5 bands cross the fermi level here.

From the results illustrated above, we find the magnetic properties
of Ba$_{1-x}$K$_x$Fe$_2$As$_2$ is very sensitive to the doping
geometry, and the magnetic moments highly depend on the ordering.
These properties imply that the magnetism of these compounds is of
itinerant character. \cite{Ba7} No matter the interlayer alignment
of the states with in-plane stripe-like AFM, they have almost the
same magnetic moments on Fe atoms. Doping does not change the nature
that the DOS near the fermi level is dominated by the FeAs layer. It
results in an increase of the N$_{E_f}$ by shifting up the bands
across the fermi level.

\section{Conclusion}
In conclusion, we have performed first-principles calculation for
Ba$_{1-x}$K$_x$Fe$_2$As$_2$ systems within the GGA+U method. Using a
negative $U_{eff}$ of $-0.50$ eV, we find the same SDW ground state
with experiment for the parent compound, and the magnetic moment on
Fe is very close to the experimental value (cal. 0.83 $\mu_B$ v.s.
exp. 0.87 $\mu_B$). We predict the most favorable doping geometries
from the energy point of view. Besides, we find that the magnetic
instability is enhanced with x=0.25, and then start to decrease.
Moreover, in our result, the magnetic structure is very sensitive to
the geometry, and the magnetic moment on Fe highly depends on
ordering, especially the in-plane ordering.

\section*{ACKNOWLEDGMENTS}
This work was partially supported by the National Natural Science
Foundation of China under Grant Nos. 20773112, 10574119, 50121202,
and 20533030, by National Key Basic Research Program under Grant No.
2006CB922004, by the USTC-HP HPC project, and by the SCCAS and
Shanghai Supercomputer Center.


\begin{thebibliography}{10}

\bibitem{jacs} Y. Kamihara, T. Watanabe, M. Hirano, and H.
Hosono, J. Am. Chem. Soc. \textbf{130}, 3296 (2008).
\bibitem{syn-Sm} X. H. Chen \textit{et al.} Nature doi: 10.1038/nature07045 (2008); arXiv0803.3603 (2008) 
\bibitem{syn-Nd} G. F. Chen \textit{et al.} arXiv:0803.4384 (2008) 
\bibitem{syn-Pr} Z. A. Ren \textit{et al.} arXiv:0803.4283 (2008) 
\bibitem{syn-Ce} G. F. Chen \textit{et al.} arXiv:0803.3790 (2008) 
\bibitem{syn-Gd} J. Yang \textit{et al.} arXiv:0804.3727 (2008) 
\bibitem{syn-Sm-hp} Z. A. Ren \textit{et al.} arXiv:0803.2053 (2008)
\bibitem{nature} Y. Kamihara \textit{et al.} Nature
\textbf{453}, 376 (2008)
\bibitem{lda1} D. J. Singh \textit{et al.} Phys. Rev. Lett.
\textbf{100}, 237003 (2008)
\bibitem{lda6} J. Dong \textit{et al.} arXiv:0803.3246 (2008)
\bibitem{ns1} C. de la Cruz \textit{et al.} arXiv:0804.0795 (2008)
\bibitem{epl} H.-H. Wen, Gang Mu, Lei Fang, Huan Yang and
Xiyu Zhu, Eur. Phys. Lett. \textbf{82}, 179009 (2008).

\bibitem{vacancy} Z. A. Ren \textit{et al.} arXiv:0804.2582 (2008)
\bibitem{Ba1} M. Rotter, M. Tegel and D. Johrendt arXiv:0805.4630
(2008)
\bibitem{Sr1} K. Sasmal \textit{et al.} arXiv:0806.1301 (2008)
\bibitem{Sr2} G. F. Chen \textit{et al.} arXiv:0806.1209 (2008)
\bibitem{Ca1} G. Wu \textit{et al.} arXiv:0806.4279 (2008)
\bibitem{Ca2} N. Ni \textit{et al.} arXiv:0806.4328 (2008)
\bibitem{Ca3} F. Ronning \textit{et al.} arXiv:0806.4599 (2008)
\bibitem{Ba2} M. Rotter \textit {et al.} arXiv:0805.4021 (2008)
\bibitem{Ba3} L. X. Yang \textit {et al.} arXiv:0806.2627 (2008)
\bibitem{Sr3} H. Y. Liu \textit{et al.} arXiv:0806.4806 (2008)
\bibitem{Ba4} F. J. Ma, Zhong-Yi Lu, and T. Xiang arXiV:0806.3526
(2008)
\bibitem{Ba5} Q. Huang \textit{et al.} arXiv:0806.2776 (2008)
\bibitem{Ba6} C. Liu \textit{et al.} arXiv:0806:3453 (2008)
\bibitem{LDA2} C. Krellner \textit {et al.} arXiv:0806.1043 (2008)
\bibitem{LDA3} I. A. Nekrasov \textit {et al.} arXiv:0806.2630
(2008)
\bibitem{neg-U} H. Nakamura \textit{et al.} arXiv:0806.4804 (2008)
\bibitem{neg-U2} H. Katayama-Yoshida \textit{et al.} arXiv:0807.3770
(2008)
\bibitem{vasp} G. Kresse and D. Joubert, Phys. Rev. B \textbf{59},
1578 (1999); G. Kresse and J. Furthmuller, Phys. Rev. B \textbf{54},
11169 (1996).
\bibitem{PAW} P. E. Bl\"{o}hl, Phys. Rev. B \textbf{50}, 17953 (1994)
\bibitem{GGA} J. P. Perdew, K. Burke, and M. Ernzerhof, Phys. Rev. Lett. \textbf{77},
3865 (1996)
\bibitem{ldau} S. L. Dudarev, G. A. Botton, S. Y. Savrasov, C. J. Humphreys and A. P. Sutton, Phys. Rev. B \textbf{57}, 1505 (1998)
\bibitem{Ba7} D. J. Singh, arXiv:0807.2643 (2008)
\end{thebibliography}
\end{document}